# Observation of polaritonic flat-band bound states in the continuum in a 2D magnet

Fuhuan Shen[1,†], Jiahao Ren[1,†], Zhiyi Yuan[2,3], Kai Wu[1], Sai Yan[1], Kunal Parasad[1], Hai Son Nguyen[2,4], Rui Su[1,3*]

1. Division of Physics and Applied Physics, School of Physical and Mathematical Sciences, Nanyang Technological University, Singapore 637371, Singapore.
2. CNRS-International-NTU-Thales Research Alliance (CINTRA), IRL 3288, Singapore 637553
3. School of Electrical and Electronic Engineering, Nanyang Technological University, Singapore, Singapore.
4. Ecole Centrale de Lyon, INSA Lyon, Université Claude Bernard Lyon 1, CPE Lyon, CNRS, INL, UMR5270, Ecully 69130, France

† These authors contribute equally

* Corresponding author. Email: surui@ntu.edu.sg

**Abstract**

Flat-band bound states in the continuum (BICs) are topological states with suppressed group velocity and robustness against radiation loss, offering a powerful platform for the exploration of non-Hermitian, nonlinear, topological phenomena and device applications. Van der Waals (vdW) metasurfaces have recently emerged as promising candidates for sustaining BICs and hybridizing with material transitions. However, the realization of flat-band BICs remains elusive. Here, we experimentally demonstrate polaritonic high-order BICs on a wide-angle flat band utilizing a subwavelength ($\sim\lambda_0/35$) metasurface made of a vdW magnet CrSBr. The large oscillator strength of direct excitons in CrSBr enables near ultrastrong coupling with BICs, leading to strongly suppressed polaritonic angular dispersions. Remarkably, second-order polaritonic BICs become flat-band across a wide angular range, with corresponding $Q$ factors exceeding 1500. Additionally, we find that these polaritonic BICs vanish in the transverse magnetic configuration, while leading to fascinating surface hyperbolic exciton-polaritons within the Reststrahlen band. Our findings underscore CrSBr as an exceptional platform for exploring flat-band photonics and polaritonics, paving the new avenue for advances in next-generation optical and quantum technologies.

**Introduction**

Photonic bound states in the continuum (BICs) are perfectly localized optical modes that exhibit no radiation leakage to the environment while coexisting with extended waves of the same energy[1-3]. Their unique properties, including inherent topological characteristics[2] and infinite $Q$ factors[4,5], drive advancements in nanophotonics and light-matter interactions[6], enabling applications in nonlinear optics[4,7-11] and chiral sources[12-14]. However, deviations from symmetric geometry or normal excitation often lead to significant degradation of $Q$ factors and energy shifts in conventional BICs[5,15,16]. Flat-band BICs address these issues, featuring near-zero group velocity and robustness to radiation loss[17-22]. Nonetheless, previous demonstrations have typically utilized passive media and relied on band edge coupling[19,20] or Moiré stacking[17,18], which limit the momentum range available for flat bands or increase modal volume and working wavelengths due to enlarged supercell units. So far, achieving strong light-matter coupling with wide-angle BIC flat bands remains highly challenging.

In parallel, Van der Waals (vdW) materials are emerging as promising building blocks for dielectric nanophotonics due to their high refractive index, wide transparency range, and intrinsic excitons, in contrast to traditional semiconductors like silicon[23-30]. As a notable example, BIC-exciton polaritons, hybrids of photonic BICs and excitons, are recently demonstrated in vdW transition metal dichalcogenides (TMDCs) metasurfaces[31,32]. The hybridization of light and matter allows highly dispersive BICs to be flattened into a more excitonic-like dispersion, though they remain far from true flat bands and can be dampened by the significant loss of intrinsic excitons[31]. Moreover, bulk TMDCs typically exhibit low quantum efficiency due to their indirect bandgap nature[33]. Recently, an emergent vdW magnet of CrSBr appears as an exceptional candidate to tackle these challenges. Unlike conventional TMDCs, it features anisotropic direct-bandgap excitons across all thicknesses[34-38]. Below the Néel temperature ($T_N$ = 132 K), due to the magnetic ordering, the excitonic oscillator strength is significantly enhanced (> 1.5 (eV)$^2$) along the magnetically easy axis (b-axis), simultaneously with their resonance sharpened (< 1 meV). These unique characteristics enable ultrastrong exciton-photon coupling achieved in CrSBr and suppress the intrinsic loss of resulting polaritons[34,35]. These advantages allow CrSBr to stand out among other TMDCs, highlighting its potential to strengthen BIC–exciton coupling and enable polaritonic BIC flat bands over a broad angular range. However, the experimental demonstration remains elusive.

Here, we experimentally demonstrate high-order polaritonic BICs on a wide-angle flat band leveraging the metasurfaces patterned from vdW magnetic CrSBr flakes. By structuring CrSBr into nanograting, we achieve polaritonic BICs approaching the ultrastrong coupling regime in the transverse electric (TE) configuration, showing strongly suppressed angular dispersions. Notably, increasing the grating's filling factor leads to the emergence of second-order polaritonic BICs, characterized by vanishing dispersion (flat band) and an enhanced $Q$ factor exceeding 1500. In the transverse magnetic (TM) configuration, these polaritonic BICs disappear, while leading to hyperbolic exciton-polaritons (HEPs) within the Reststrahlen (RS) band, where the b-axis permittivity is negative. These HEPs exhibit angular-invariant dispersion in both reflectance and photoluminescence (PL) measurements, attributed to the deeply subwavelength confinement of the surface mode. Our findings highlight CrSBr's unique ability to intrinsically tune metasurface photonic properties via its excitons, paving the way for advanced photonic and polaritonic devices in quantum circuits and magneto-optic applications.

**Results**

**Suppressed angular dispersion of polaritonic BICs in CrSBr gratings**

CrSBr is the layered vdW materials having an orthorhombic structure with space group *Pmmn* ($D_{2h}$). Below the bulk Néel temperature (~132 K), CrSBr exhibits the A-type antiferromagnetic order with the magnetic moments oriented ferromagnetically in b-axis within the ab-plane while ordered antiferromagnetically in the c-axis (see top panel in Fig. **1a**). The electronic and structural anisotropy, along with the magnet-correlated properties, result in the tightly bound quasi-1D excitons along the b-axis (magnetic easy axis)[37]. This behaviour is clearly reflected in the permittivities measured in different directions (Fig. **1b**). Along the b-axis, the dominant contribution to the excitonic dispersion comes from the main exciton (X), which has an energy of approximately 1.3655 eV, an oscillator strength of $f_X = 1.6$ $(eV)^2$, and a linewidth of $\gamma_X = 0.85$ meV, while the influence of side excitons (e.g., X*) is negligible (see **Methods**). By contrast, the permittivity along a-axis ($\varepsilon_a$) has no contributions from excitons.

Here, the exfoliated CrSBr flakes (from ~15 nm to ~35 nm) are patterned into the nanograting metasurfaces (as schematically illustrated in the bottom panel of Fig. **1a**). The TE modes are first studied where the electric field (**E**) is along the long axis (defined as *y*-axis) of grating bar. The b-axis of CrSBr also aligns with the electric field TE configuration (Fig. **1c**). The periodic structure (i.e., nanograting) introduces the diffractive coupling between two propagating waves within the CrSBr slab, resulting in the splitting of two photonic modes, with one being the bright mode, i.e., guided mode resonance (GMR), and another being the dark mode, namely BIC mode[39,40] (also see **Supplementary Section A**). When main excitons are artificially "turn off" ($f_X$=0 $(eV)^2$), these two photonic modes (indicated by a blue and a red dashed line in Fig. **1d**, respectively) appears in a significantly higher energy (~ 2.55 eV) above the original excitonic resonance, leading to a substantial detuning Δ (> 1 eV) between the photonic modes and the excitons. After "turning on" the main excitons ($f_X$ =1.6 $(eV)^2$), the GMR/BIC modes independently hybridize with the excitons, resulting in the formation of GMR/BIC upper ($UP_{GMR}/UP_{BIC}$) and lower polaritons ($LP_{GMR}/LP_{BIC}$), which can be described by the Hamiltonian:

$$H = H_{cav} + H_{exc} + g(\hat{a}\sigma_+ + \hat{a}^\dagger\sigma_+ + h.c.) + \frac{g^2}{\omega_a}(\hat{a}^\dagger + \hat{a})^2. \quad (1.)$$

The first two terms $H_{cav} = \omega_c(\frac{1}{2} + \hat{a}^\dagger\hat{a})$ and $H_{exc} = \omega_a(\frac{1}{2} + \sigma_z)$ represent the energy of uncoupled photonic modes (i.e., GMR or BIC modes) and excitons. The third and fourth term indicate the photon-exciton interaction. As the system approaches the ultrastrong coupling regime (see **Supplementary** Fig. **S2** where the coupling strength $g = 118$ meV is close to $0.1\omega_a$), the fast-rotation term ($\hat{a}^\dagger \sigma_+$) will be kept and so-called $A^2$ term arise as the last term in Hamiltonian[41-43]. Due to the large detuning which is defined as energy difference of photonic mode and exciton, i.e., $\Delta = \omega_c - \omega_a$, the resultant upper polaritons with dominant photonic weight exhibit a slight blue shift with respect to the original photonic modes, but with a similar angular dispersion with the corresponding photonic modes. In contrast, the lower polaritons show significantly suppressed angular dispersion, owing to dominant excitonic weight (details can be found in **Supplementary Section B**).

The angle-resolved reflectance spectra calculated by rigorous coupled-wave analysis (RCWA) provide clearer insights into distinct behaviours of GMR- and BIC-polaritons (Fig. **1e**). The upper GMR-polariton (UP$_{GMR}$) exhibits a very broad linewidth, while the linewidth is significantly reduced for the lower GMR-polariton (LP$_{GMR}$) due to the dominant contributions from the main excitons, which have a narrow linewidth of 0.85 meV. By contrast, UP$_{BIC}$ manifested itself as the vanishing mode as it approaches the normal direction (at-Γ point), whose linewidth becomes broaden with the increasing incident angle. The LP$_{BIC}$ shows a similar trend, but with a reduced linewidth and strongly suppressed angular dispersion.

The corresponding calculated near-field distributions (Fig. **1f**) reveal different instincts of GMR and BIC polaritons, respectively. GMR-polaritons exhibit the symmetric near field distributions (first top two images of Fig. **1f**), thus being able to couple to the radiation in the far field. While by striking contrast, BIC polaritons show the antisymmetric field distributions, preventing the coupling with the radiative continuum due to the parity mismatch. Additionally, compared to the upper polaritons, the lower counterparts demonstrate greater field confinement as evidenced by the field distribution profiles at the center of grating bars (right panels in Fig. **1f**) where the dashed lines indicated the boundary of CrSBr grating bars.

**Experimental demonstration of BIC-polaritons in CrSBr nanogratings**

To fabricate the CrSBr nanogratings (Fig. **2a**), the CrSBr flake is first exfoliated onto a PDMS substrate and then transferred onto a SiO$_2$ (300 nm)/Si substrate that has been previously evaporated with Cr markers for the location of materials (i). This is then followed by a standard electron-beam lithography (EBL) process to create the top layer nanograting pattern made of ZEP photoresist (ii). Finally, the CrSBr flake is etched using the ZEP nanopattern as a mask with the inductively coupled plasma (ICP) dry etching process (iii). The ZEP residue is subsequently removed by the O$_2$/Ar plasma (see more details in **Methods**). Compared to the previous works[31,44], the simplified fabrication procedures adopted here don't require a hard mask made of metal or silicon nitride, leading to the reduction of residues on the CrSBr surface. Figure **2b** exhibits the microscopic image of a representative sample with the thickness of pristine CrSBr of around 26 nm. The periods of these three gratings are fixed at 400 nm but with various filling factors (defined by the width of grating bar divided by the period, as schematically shown in Fig. **1c**). The scanning electron microscope (SEM) image (bottom image in Fig. **2c**) of one of representative gratings shows a filling factor of around 0.35. The atomic force microscopy (AFM) profile exhibits the same thickness of grating bars as that of the pristine layer, indicating the complete etching of the grating structure.

The angle-resolved differential reflectance spectra reveal completely distinct optical behaviour of the CrSBr grating, compared with that of pristine CrSBr flake (Fig. **2d**). While only the signatures of main (X) and side ($X^*$) excitons are unambiguously shown for the pristine CrSBr flake, the $LP_{GMR}$ and $LP_{BIC}$ branches are emergent for the CrSBr grating, same as those shown in Fig. **1e**. Notably, there is an additional polaritonic band, marked by a black arrow, appearing below the X energy with a small but unambiguous red shift. In the following analyses, we show that this is actually not a single band but an assembly of multiple higher-order lower polariton states. All above polariton branches disappear when the electric field is oriented perpendicular to the b-axis (left panel in Fig. **2d**) due to the quasi-1D nature of excitons.

The $LP_{GMR}$ and $LP_{BIC}$ with the significantly suppressed angular dispersions are also manifested in the PL measurements (Fig. **2e**). The extracted resonant frequency $\omega_0$ of $LP_{GMR}$ (Fig. **2f**) is nearly independent on the in-plane momentum $k_x$, while $\omega_0$ of $LP_{BIC}$ shows a slightly red shift (~ 1 meV) as $k_x$ increases from 1 μm$^{-1}$ to 3 μm$^{-1}$. The $Q$ factor (Fig. **2g**) of $LP_{GMR}$ approaches 700 and shows slight increase with the increasing $k_x$. By contrast, a significant decrease in the $Q$ factor of $LP_{BIC}$ is observed, dropping from around 700 at $k_x=1$ μm$^{-1}$ to around 400 at $k_x=3$ μm$^{-1}$ due to the symmetry breaking induced by non-zero $k_x$. In addition, reflectance spectra at different $k_y$ (with $k_y=0, 2.3, 3.5$ μm$^{-1}$) indicates the resonance frequency and linewidth of $LP_{GMR}$ and $LP_{BIC}$ are almost invariant in $k_y$ direction (Fig. **2h**).

**Polaritonic flat-band high-order BICs**

As aforementioned, the polaritonic band, appearing slightly below the X energy in Figure **2d**, is ascribed to the assembly of high-order polaritons. As schematically shown in Figure **3a,** the photonic GMR-BIC pair of $n$-th order ($n=1, 2, 3 \ldots$) appear sequentially above the X, corresponding to the folding and coupling of the same TE guided modes but operating at wavevector $\pm n\frac{2\pi}{P}$. Their hybridizations with X result in the first, second, and higher order LP branches, which approach X with the increasing order. Through tuning the geometric parameters such as period or filling factor (see **Supplementary** Fig. **S3**), the second- or even third-order polaritons can be separated. When the filling factor is increased to 0.5 (noted by the green triangle in **Fig. 2b**, whose SEM image can be seen in **Supplementary** Fig. **S4a**), the first-order LP$_{BIC}$ and LP$_{GMR}$ show an unambiguous red shift and become more angularly dispersive due to the increased photonic contributions in resulting polaritons (**Supplementary** Fig. **S6**), particularly for LP$_{BIC}$. More interestingly, additional pair of LP$_{BIC}$ and LP$_{GMR}$ emerges between the first-order polaritons and X (**Supplementary** Fig. **S4b-c**). As the filing factor further increased to ~0.72, as shown in the SME image in Figure **3c**, the first-order LP$_{BIC}$ and LP$_{GMR}$ show a pronounced red shift and become more dispersive (Fig. **3d**). The zoom-in reflectance and PL spectra (Fig. **3e**) exhibit that in addition to the second-order LP$_{BIC}$ (noted by a blue arrow) and LP$_{GMR}$ (noted by a red arrow), the third-order LP$_{GMR}$ is emergent (noted by a green arrow). The corresponding calculated near field distributions confirm their high-order nature where the node (i.e., $E_y$=0) number is increased from 2 (first-order LP$_{GMR}$) to 4 and 6 for second- and third-order LP$_{GMR}$. For second-order LP$_{BIC}$, the node number is increased from 1 (first-order LP$_{BIC}$) to 3. We noted that although in simulated results (left panel in Fig. **3d**), the third-order LP$_{BIC}$ appears slightly below the third-order LP$_{GMR}$, it cannot be distinguished in the measured results due to the narrow energy separation and low optical contrast.

The extracted resonant frequencies of second-order LP$_{GMR}$ and LP$_{BIC}$, and the third-order LP$_{GMR}$ exhibit nearly vanishing angular dispersions (<0.2 meV·µm), towards the perfect polaritonic flat bands (Fig. **3f**). The third-order LP$_{GMR}$ exhibits the $Q$-factor of approximately 1300 to 1400, while the second-order LP$_{BIC}$ shows a $Q$-factor exceeding 1500, which is, to the best of our knowledge, the record value for the vdW photonics in the near-infrared range[31]. The upper limit of $Q$-factor is set by the intrinsic linewidth (0.85 meV) of excitons, giving the largest Q factor of around 1600 calculated by the exciton energy divided by the linewidth. Although the $Q$ factor decreases with increasing $k_x$, it still maintains above 1000 at $k_x$=3µm$^{-1}$. The angle-resolved reflectance spectra (Fig. **3h**) at different $k_y$ also indicates the higher-order polariton modes is independent on momentums in arbitrary directions. Similar trends can also be found in the CrSBr grating with ff=0.5 (**Supplementary** Fig. **S4g**). Due to the limitation of collection angle of objective (NA = 0.65) the measured range is around 40 degrees, while the calculation indicates an almost 90-degree wide-angle flat band of high-order BICs (**Supplementary** Fig. **S7**).

**Hyperbolic exciton-polaritons at TM configurations**

As indicated in Figure **1b**, CrSBr is an anisotropic material with different in-plane permittivities. Re($\varepsilon_a$)>0 is across all the energy range. Below the X energy, Re($\varepsilon_b$)> 0 while Re($\varepsilon_b$) ≠Re($\varepsilon_a$), resulting in the polaritons with the elliptical isofrequency contours (left part of Fig. **4a**), as analyzed above in the TE configuration. By contrast, above the X energy, the signs of permittivities along a- and b-axis (Re($\varepsilon_b$)< 0) become opposite, yielding excitonic Reststrahlen (RS) band[45]. This leads to the so-called hyperbolic exciton-polaritons (HEPs) due to the hyperbolic isofrequency contours[46] (right part of Fig. **4a**). HEPs are TM propagating modes existing below the light line, showing subdiffractional wavelength confinement and enhanced light-matter interaction compared with traditional exciton-polaritons. Consequently, these HEPs, which are difficult to probe in the far field, can usually be accessed only through near-field spectroscopy, such as scattering-type scanning near-field optical microscope (s-SNOM)[46]. Here, we overcome this limitation by employing a nanograting structure to fold the HEPs at $\frac{2\pi}{P}$ into $k_x=0$ (Fig. 4d), which allows direct far-field excitation and detection. Similar to probe the surface plasmon polaritons with grating coupler[47], the original surface mode become Bloch HEPs by our periodic structures (as the near field distributions in Fig. **4d** indicates)[46].

To excite Bloch HEPs in experiments, the system is set as the TM configuration where the electric (magnetic) field is perpendicular (parallel) to the long axis of grating bar, with the b-axis of CrSBr aligned with the electric field (as schematically shown in Fig. **4b**). The period $P=400$ nm is chosen to match the HEP propagating wavevector range (around 15-20 $\mu m^{-1}$) of CrSBr[46]. Measured reflectance spectra at normal direction (Fig. **4c**) exhibits these HEPs above the X energy (Fig. **4c**), with the corresponding PL measurements showing the consistent results (Fig. **4c**). The unambiguous redshift with filling factor is ascribed to the higher average permittivity of CrSBr slab. Angle-resolved reflectance (Fig. **4e**) and PL (Fig. **4f**) spectra indicate the flat-band behaviors of HEPs in the momentum space which can be ascribed to their deeply subwavelength-confined mode volumes[22]. Different from the conditions where the GMR and BIC polaritons are emergent in the TE configuration, there are no corresponding counterparts for TM configuration. As for the formation of TM modes, it requires field retardation along the *z*-direction (i.e., c-axis of CrSBr). However, the subwavelength thickness (< 30 nm) and low permittivity ($\varepsilon_c=4$ at 4K) are significantly inadequate to support the TM guided modes[48].

**Discussion**

Our work presents the first demonstration of polaritonic BIC flat bands in deeply subwavelength CrSBr nanogratings. The angular-dispersions of these (lower) BIC-polaritons are significantly suppressed due to the ultrastrong coupling and large detuning between photonic BICs and excitons in CrSBr. The increasing of the filling factor of the grating enables the observation of second-order BIC-polaritons, showing vanishing angular dispersion and enhanced and robust $Q$ factors. The flat-band BIC-polaritons achieved in this work exhibits the exceptional robustness of resonant energies and robust $Q$ factors against the inclined incidence, which outperformance the traditional dispersive BICs and the flat-band BICs achieved through the hybridization of different energy bands. In addition, in the TM configuration, the surface HEPs are observed within the RS band by reflectance and PL characterizations, showing the angular-invariant dispersion behaviour.

Our findings highlight the significant potential of the vdW magnet CrSBr for advanced applications in nanophotonics and polaritonics. Recently, pioneering studies have explored the nonlinear properties and polariton condensates in CrSBr; however, these efforts have been limited by the bulky Fabry-Pérot cavity structures, which would hinder reductions in device footprint and compatibility with on-chip applications for integrated photonics and quantum sources. The deliberate designs of CrSBr structures overcome these limits and allow for the giant flexibility to manipulate the novel optical properties such as light wavefronts and topological behaviours, paving the new avenue for the interplay of the photonic, electronic, and magnetic properties in this promising flatform.

**Methods**

**Device fabrications**

The pristine CrSBr flakes are prepared on PDMS by mechanical exfoliation from the CrSBr crystal bought from HQ graphene. The CrSBr flakes with homogeneous and target thickness (10 nm to 40 nm,) are selected according to the microscope images. Then, using the dry-transfer method, the selected CrSBr flake is transferred onto the targeted area marked by the Cr makers. Next, the CrSBr flake is cleaned by the $O_2$/Ar plasma to remove the residues on the surface and then is spin-coated with ZEP (pristine ZEP is diluted by the anisole by the volume ratio of 1:1) photoresist at 5000 rpm for 1 min. A standard EBL is applied and followed by a development process (immersed in pentyl acetate for 1 min at zero degree and rinsed in isopropyl alcohol for 30s at room temperature). Last, the CrSBr flake with top layer nanopatterns is etched by the ICP process. After ICP, the residual ZEP on the surface of CrSBr can be removed by the Ar/$O_2$ plasma.

**Fitting parameters**

The permittivity of CrSBr along b-axis were extracted by fitting the measured reflection spectra of unpattern CrSBr area with a multi-Lorentzian model:

$$H\varepsilon_b(\omega) = \varepsilon_{b0} + \sum_j \frac{f_j}{\omega_j^2 - \omega^2 + i\omega\gamma_j}, \qquad (2.)$$

where $\varepsilon_{b0} = 11$ is the background permittivity, $f_j$, $\omega_j$, and $\gamma_j$ are the oscillator strength, resonant frequency, and linewidth of the exciton respectively. The fitting parameters are listed below:

X: $f_1$=1.6 (eV)$^2$, $\omega_1$=1.3655 eV, and $\gamma_1$ = 0.85 meV.

X$^*$: $f_2$= 0.2 (eV)$^2$, $\omega_2$=1.3801 eV, and $\gamma_2$ = 5 meV.

X$^{**}$: $f_3$= 0.025(eV)$^2$, $\omega_3$=1.3908 eV, and $\gamma_3$ = 3 meV.

X$_2$: $f_4$=1.1 (eV)$^2$, $\omega_4$ = 1.76 eV, and $\gamma_4$ = 23 meV.

X$_2$, representing the B exciton residing around 1.76 eV, though has a large oscillator strength compared with other side excitons, which is far detuned from X and thus has negligible influence on the lower polaritons near X. Here, the $\varepsilon_a = 11$ and $\varepsilon_c = 4$ are adopted in our simulations.

**Isofrequency contours**

The isofrequency contours can be described by

$$\frac{k_x^2}{\varepsilon_y} + \frac{k_y^2}{\varepsilon_x} = k_0^2, \tag{3.}$$

where out-of-plane $k_z$=0 since we focus on the in-plane properties by propagating modes. $k_0$ is the free-space wave vector, and $\varepsilon_x$ ($\varepsilon_y$) is the x- (y-) component of dielectric tensor, corresponding to $\varepsilon_a$ ($\varepsilon_b$) in Fig. **4a**.

**Numerical simulations**

The angle-resolved reflectance spectra are calculated with the RCWA method based on the permittivities mentioned above. To match the dielectric environment in experiment, the refractive index is set as *n*=1 for the top layer (air) and *n*=1.46 for the bottom layer (SiO$_2$). For the near field calculation, the eigen frequency of GMR-polariton mode at normal direction is first confirmed by the reflection spectrum and then the near field is calculated at the choose frequency for this mode. For the BIC-polariton mode, because it is strictly dark at normal direction, we thus calculated the spectrum at near-normal (0.01°) and then performed the near field calculation.

**Optical measurements**

The angle-resolved measurements were based on the home-built Fourier transform setup (see Supplementary Fig. **S8**). The signal was collected by an objective of NA=0.65. For the reflectance, the differential reflectance $\frac{\Delta R}{R}$, defined by ($R_{\text{sample}} - R_{\text{sub}}$)/$R_{\text{sub}}$ where $R_{\text{sample}}$ ($R_{\text{sub}}$) represents the reflectance by the sample (substrate), respectively, are measured for both the pristine CrSBr flake and nanogratings. The white light was applied as the excitation field whose polarization was aligned with the b-axis of CrSBr b a polarizer. For the PL measurement, the laser with wavelength of 532 nm impinged on the sample and the PL signal was collected.

**Data availability**

All data that supports the plots within this paper and other findings of this study are available from the corresponding author upon request.

**Acknowledgements**

R.S. gratefully acknowledge funding support from the Singapore Ministry of Education via the AcRF Tier 2 grant (MOE-T2EP50222-0008), AcRF Tier 3 grant (MOE-MOET32023-0003) "Quantum Geometric Advantage" and Tier 1 grant (RG90/25). R.S. also gratefully acknowledges funding supports from Nanyang Technological University via a Nanyang Assistant Professorship start-up grant and the Singapore National Research Foundation via a Competitive Research Program (grant no. NRF-CRP23-2019-0007). H.S.N. acknowledge funding support from the French National Research Agency (ANR) under the project POLAROID (ANR-24-CE24-7616-01).

**Author contributions**

F.S. and R.S. conceived the idea. F.S., K.W., S.Y., and K.P. fabricated the samples. J.R, F.S., and S.Y. performed the optical experiments. F.S. and Z.Y. made the simulations. All the authors contributed to the data analyses and paper writing. R.S. supervised the research.

**Competing interests**

The authors declare no competing interests.

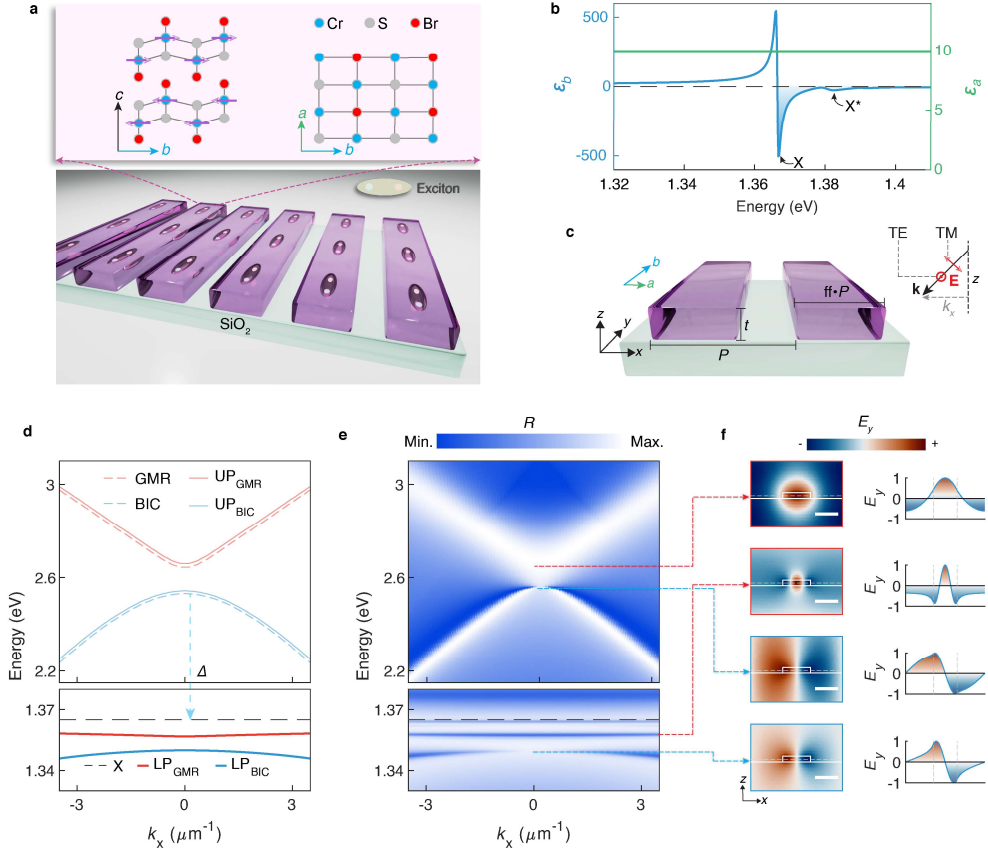

**Fig. 1| Suppressed angular dispersions of BIC-polaritons in CrSBr gratings.** (**a**) Schematic illustration of CrSBr nanogratings (bottom). The blue and red circles represent the electron and hole forming the exciton respectively. Top panel shows the atomic arrangements of CrSBr where the purple arrows indicate the magnetic moment. (**b**) Real parts of permittivities of CrSBr along b-axis (blue) and a-axis (green) respectively. (**c**) Definition of geometric parameters of CrSBr grating. $P$ represents the period; $t$ represents the thickness; and filling factor (ff) is defined as the width ($w$) of grating bar divided by $P$. The b-axis of CrSBr is along the long axis of grating bar ($y$-axis). The $x$-axis ($y$-axis) is defined as the direction perpendicular (parallel) to the long axis of grating bar. The TE configuration is set with the electric field parallel to the long axis of grating bar, while for TM configuration, the electric field is within the incident plane where the magnetic field is perpendicular to the long axis of grating bar. (**d-e**) Calculated energy dispersions (**d**) of uncoupled GMR and BIC modes, $UP_{GMR}$, $UP_{BIC}$, $LP_{GMR}$, $LP_{BIC}$, and X respectively and corresponding angle-resolved spectra (**e**) of the CrSBr grating calculated by RCWA method. (**f**) Left panel: near field ($E_y$) distributions of $UP_{GMR}$, $LP_{GMR}$, $UP_{BIC}$, and $LP_{BIC}$ in $x$-$z$ profile. Right panel: corresponding amplitude distributions of $E_y$ indicated by the dash lines in left panel. The gray dashed lines indicate the width of grating bar. For the calculations in (**d-f**), $P$=400 nm, $t$=25 nm, and ff=0.3 are adopted.

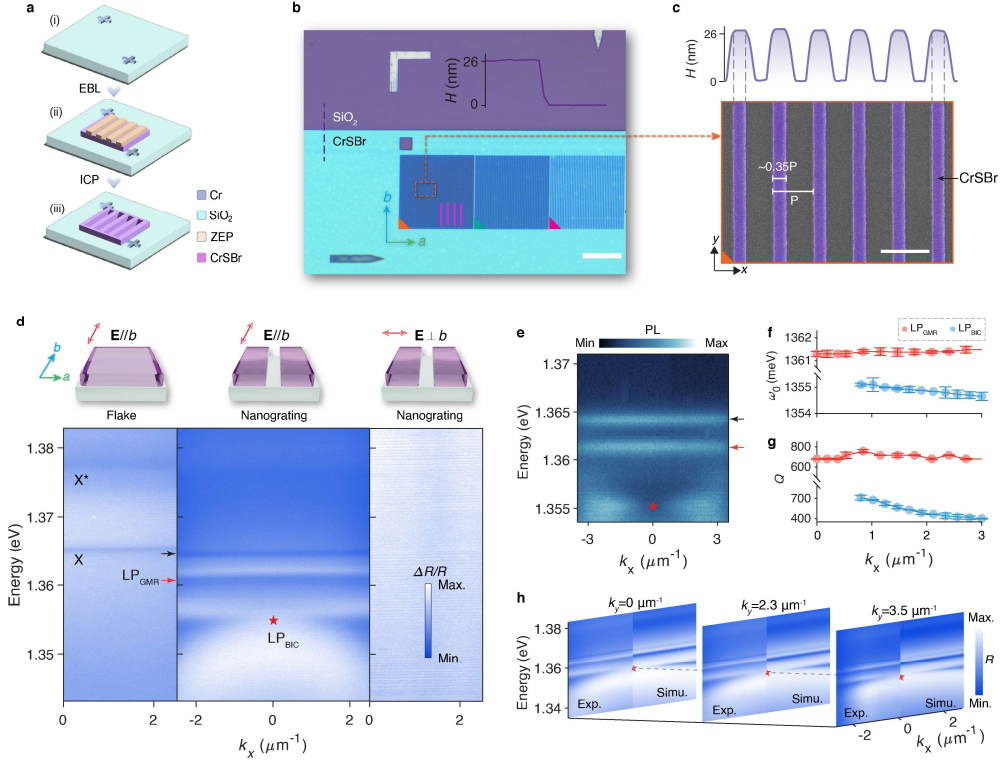

**Fig. 2| Experimental demonstration of BIC-polaritons in CrSBr nanogratings.** (**a**) Schematic fabrication procedures of CrSBr metasurfaces. (**b**) Microscopic image of CrSBr nanogratings. The inset shows the AFM height profile of the CrSBr flake across the boundary which is indicated by the dashed-purple line. The scale bar represents 10 μm. The direction of grating bar is indicated by the schematic image. (**c**) Bottom panel: SEM image (bottom) of a CrSBr grating with filling factor of ~0.35. The fake color (purple) is adopted to emphasize the CrSBr grating bars. The scale bar represents 500 nm. Top panel shows the corresponding AFM height profile of the grating. (**d**) Angle-resolved reflection spectra of bulk CrSBr flake (left), the CrSBr nanograting with excitation electric field along the b-axis (middle) and perpendicular to the b-axis (right), respectively. Black and red arrows indicate the (condensed) high order polaritons and LP$_{GMR}$, respectively. The red star indicates vanishing signature at $k_x = 0$ for LP$_{BIC}$. (**e**) Corresponding angle-resolved PL spectra. (**f-g**) Resonant frequencies (**g**) and $Q$ factors (**h**) for LP$_{GMR}$ (red solid circles) and LP$_{BIC}$ (blue solid circles). Error bars arise from the Lorentzian fittings. (**h**) Angle-resolved reflection spectra of CrSBr grating with $k_y$ varied from 0, 2.3, to 3.5 μm$^{-1}$. All the angle-resolved measurements are along the $k_x$ direction. Unless otherwise stated, all tests are conducted at the 4K temperature.

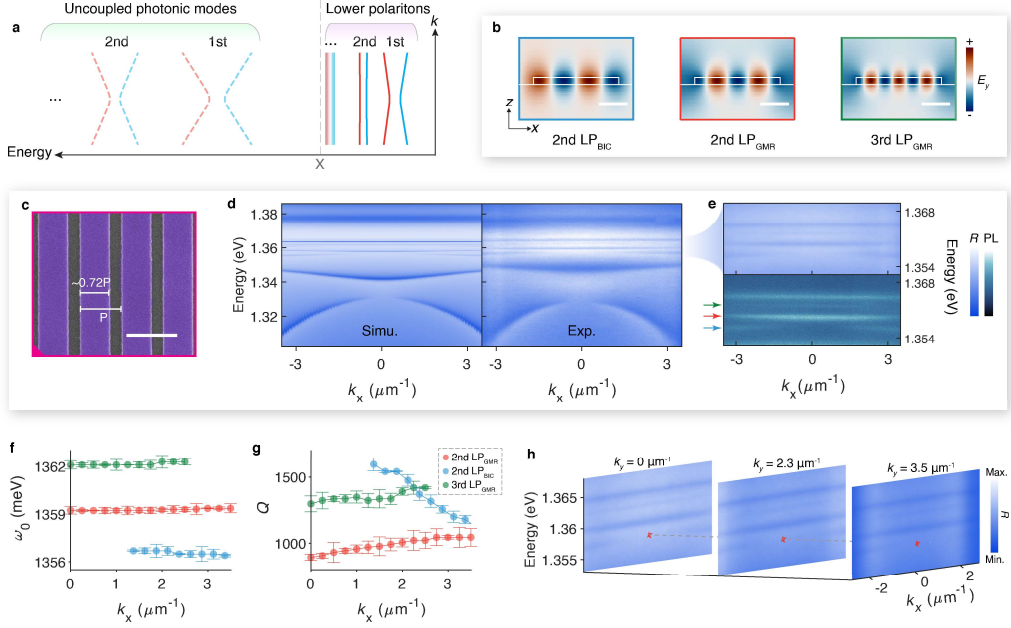

**Fig. 3| High-order BIC and GMR polariton modes.** (**a**) Schematic formation of high order $LP_{BIC}$ and $LP_{GMR}$. (**b**) Calculated near-field ($E_y$) distributions for the second-order $LP_{GMR}$, $LP_{BIC}$, and third-order $LP_{GMR}$, respectively. (**c**) SEM image of the CrSBr grating with a filling factor of around 0.72. (**d**) Simulated and measured angle-resolved reflectance spectra of the grating shown in (**c**). (**e**) Corresponding zoom-in reflectance and PL spectra shown in (**d**). (**f-g**) Extracted resonant frequencies $\omega_0$ (**f**) and $Q$ factors (**g**) of second-order $LP_{GMR}$ (red circles), second-order $LP_{BIC}$ (blue circles), and third-order $LP_{GMR}$ from the PL spectra shown in (**e**). (**h**) Corresponding angle-resolved reflectance spectra with $k_y$ varied from 0, 2.3, to 3.5 $\mu m^{-1}$.

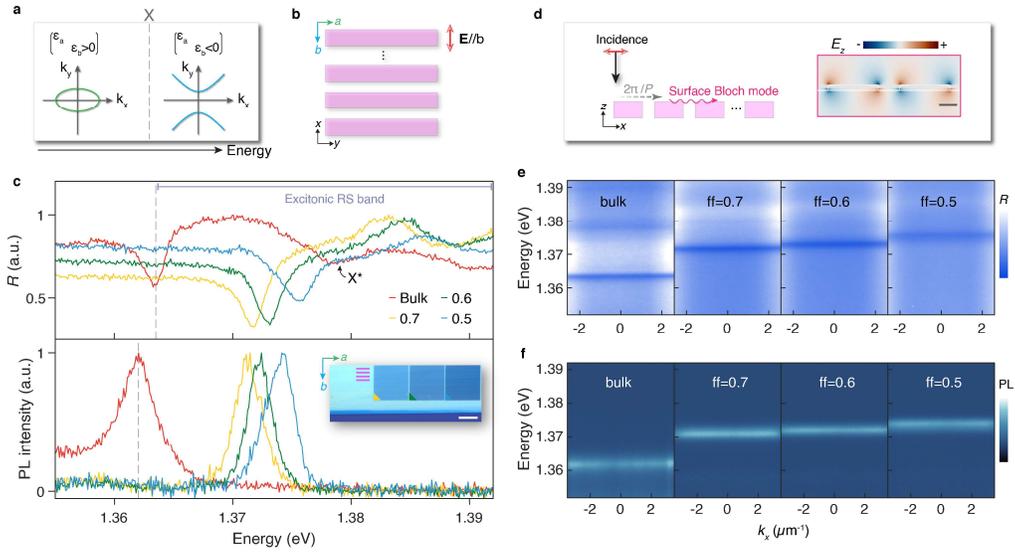

**Fig. 4| Hyperbolic polaritons in excitonic RS band.** (**a**) Schematic isofrequency contours for $\varepsilon_b > 0$ and $\varepsilon_b < 0$ which is separated by excitonic resonance (grey dashed line). (**b**) Microscopic image of CrSBr grating at TM configuration. The scale bar represents 10 µm. (**c**) Normal reflectance (top panel) and PL (bottom panel) spectra for bulk (red curve) and gratings with different filling factors (ff=0.7: yellow curve; ff=0.6: green curve; ff=0.5: blue curve). The slight difference between reflectance dip and PL peak of bulk CrSBr is ascribed to the Stokes shift. (**d**) Calculated near field distributions of HEPs. Scale bar represents 100 nm. (**e-f**) Measured angle-resolved reflectance (**e**) and PL (**f**) spectra.